\documentclass{article}

\usepackage{spconf}
\usepackage[T1]{fontenc}
\usepackage[utf8]{inputenc}
\usepackage{newtxtext,newtxmath}
\usepackage{amsmath,mathtools}
\numberwithin{equation}{section}
\usepackage{graphicx}
\usepackage[labelformat=simple, labelsep=period]{caption}
\usepackage{subcaption}
\usepackage{algorithm}
\usepackage{algpseudocode}
\algrenewcommand\algorithmiccomment[1]{\hfill\(\triangleright\) #1}
\usepackage{microtype}
\usepackage[hidelinks]{hyperref}

\title{Prompt-aware classifier free guidance for diffusion models}
\name{Xuanhao Zhang$^{1}$ \qquad Chang Li$^{2}$\sthanks{Corresponding author. Email: lc\_lca@mail.ustc.edu.cn}}
\address{$^{1}$ China Pharmaceutical University \\
         $^{2}$ University of Science and Technology of China}
\begin{document}
%
\maketitle
\begin{abstract}
Diffusion models have achieved remarkable progress in image and audio generation, largely due to Classifier-Free Guidance. However, the choice of guidance scale remains underexplored: a fixed scale often fails to generalize across prompts of varying complexity, leading to oversaturation or weak alignment. We address this gap by introducing a prompt-aware framework that predicts scale-dependent quality and selects the optimal guidance at inference. Specifically, we construct a large synthetic dataset by generating samples under multiple scales and scoring them with reliable evaluation metrics. A lightweight predictor, conditioned on semantic embeddings and linguistic complexity, estimates multi-metric quality curves and determines the best scale via a utility function with regularization. Experiments on MSCOCO~2014 and AudioCaps show consistent improvements over vanilla CFG, enhancing fidelity, alignment, and perceptual preference. This work demonstrates that prompt-aware scale selection provides an effective, training-free enhancement for pretrained diffusion backbones.
\end{abstract}
\section{Introduction}
\label{sec:intro}

Diffusion model~\cite{ho2020denoising} have achieved remarkable success in generative tasks such as image~\cite{saharia2022photorealistic}, speech~\cite{kong2020diffwave}, and audio~\cite{liu2023audioldm2, li2024quality} synthesis, largely due to their stable training and ability to produce high-fidelity samples. To further improve controllability, CFG~\cite{ho2022classifier} has become the dominant technique, where conditional and unconditional predictions are combined to guide the sampling trajectory. By scaling the difference between these two predictions, CFG can greatly enhance alignment with conditioning signals and improve corresponding generation quality in the meantime, leading to substantial improvements in tasks for content generation. Owing to its simplicity and effectiveness, CFG has now become a standard component in diffusion-based generative modeling.

Although diffusion models with proper guidance strategies can already approximate the score of the data distribution and yield strong sampling performance, there remain several opportunities to push their limits, especially for pretrained diffusion backbones. One line of research explores structural inductive biases inside the network to improve sampling without retraining, such as FreeU~\cite{si2024freeu} and MelRefine~\cite{guo2024mel}. Another line focuses on training-free guidance refinements beyond CFG, including Autoguidance~\cite{karras2024guiding, kasymov2024autolora}, TPG~\cite{rajabi2025token}, and SAG~\cite{hong2023improving}, which alleviate the inherent limitations of the unconditional branch in CFG. In addition, a series of works investigate fundamental components of the sampling process itself: Goldnoise~\cite{zhou2024golden} examines the role of noise initialization, while CATImage~\cite{li2025cost} adaptively chooses sampling timesteps based on the type of conditioning signal, aiming for a more optimal trade-off between quality and efficiency. These advances collectively suggest that even for pretrained diffusion models, there is still room to improve controllability, stability, and quality in a training-free manner.

\begin{figure}[t]
    \centering
    \begin{subfigure}[t]{0.18\textwidth}
        \centering
        \includegraphics[width=\textwidth]{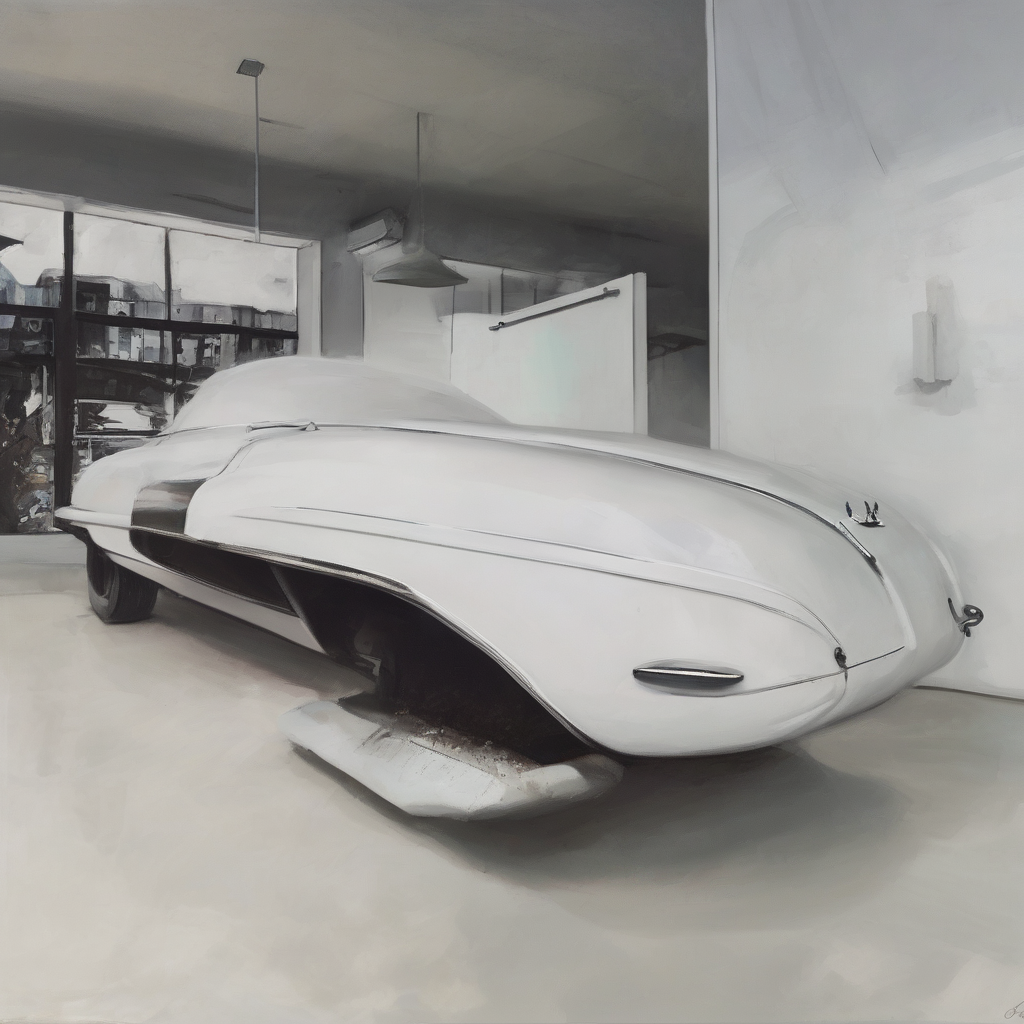}
        \caption*{(a) $w_\text{default} = 5.0$}
    \end{subfigure}
    \begin{subfigure}[t]{0.18\textwidth}
        \centering
        \includegraphics[width=\textwidth]{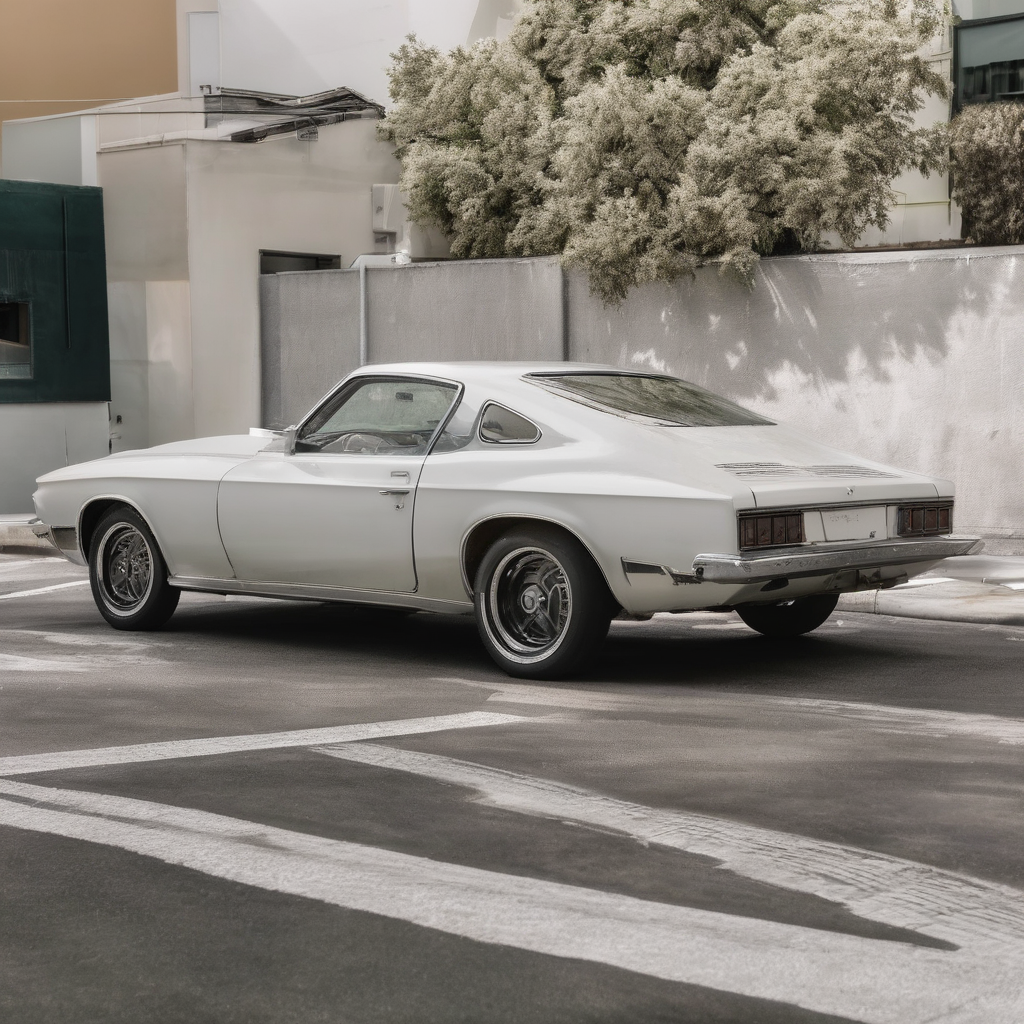}
        \caption*{(b) $w_\text{optimal} = 3.0$}
    \end{subfigure}
    \caption{A white car.}
    \label{fig:scale1}

    \begin{subfigure}[t]{0.18\textwidth}
        \centering
        \includegraphics[width=\textwidth]{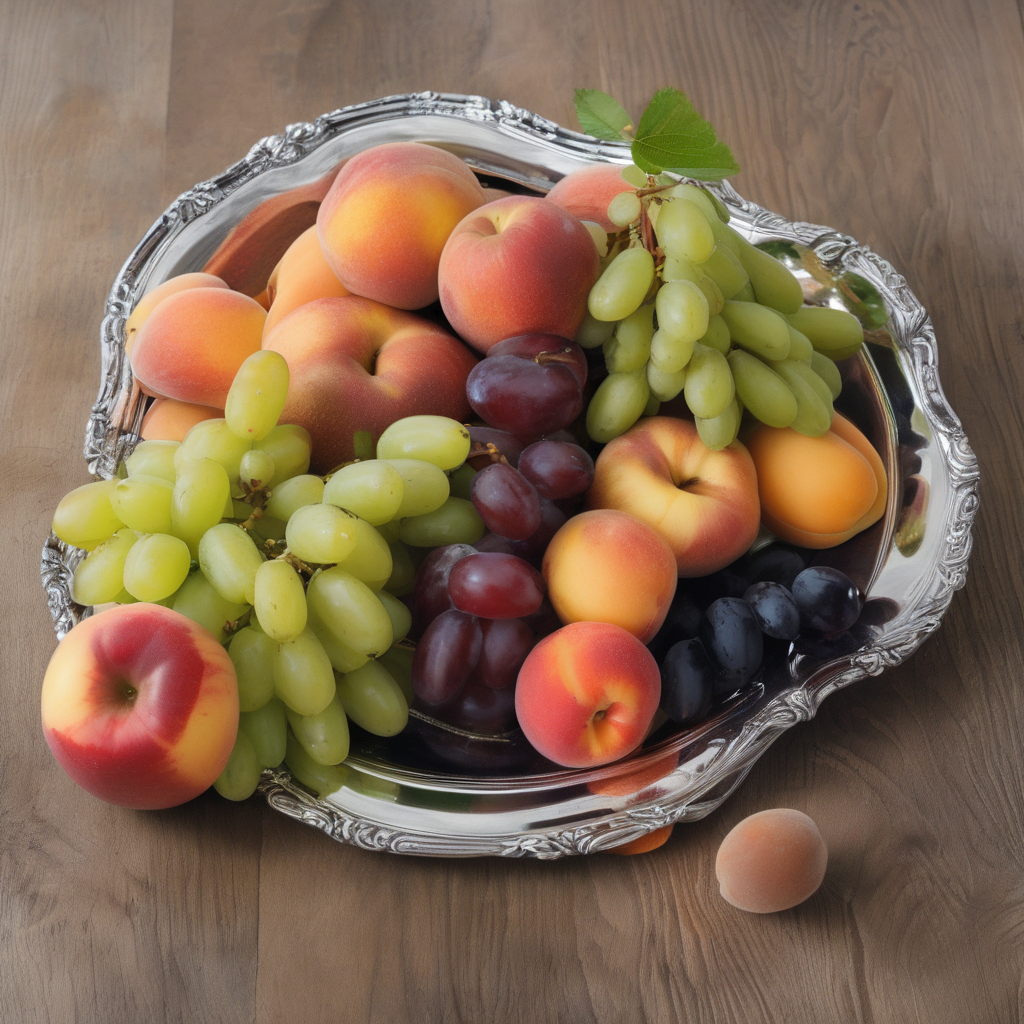}
        \caption*{(a) $w_\text{default} = 5.0$}
    \end{subfigure}
    \begin{subfigure}[t]{0.18\textwidth}
        \centering
        \includegraphics[width=\textwidth]{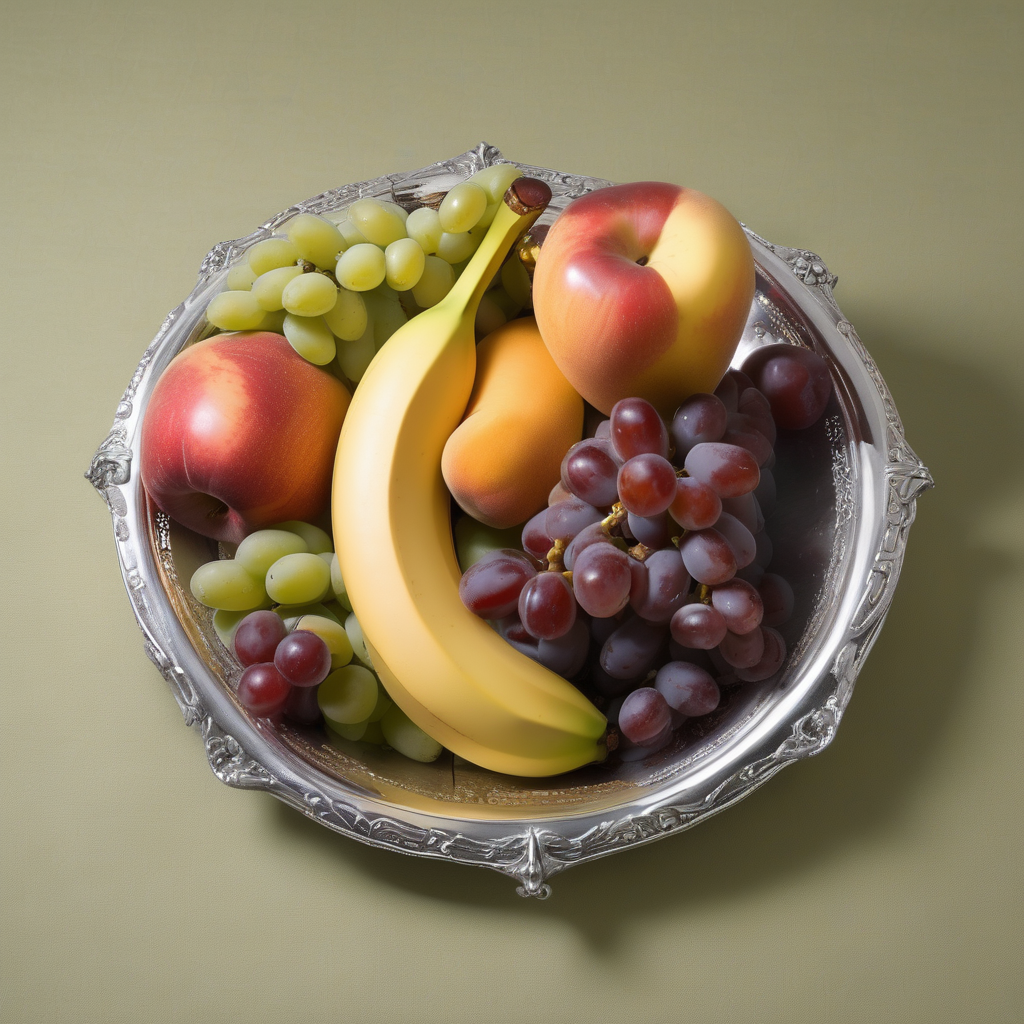}
        \caption*{(b) $w_\text{optimal} = 8.0$}
    \end{subfigure}
    \caption{A delicious fruit plate in a silver bowl. The fruit includes an apple, two apricots, red and green grapes, a banana, and a peach.}
    \label{fig:scale2}
\end{figure}

However, the selection of the guidance scale, despite being one of the most basic hyperparameters in CFG, has received surprisingly little attention. As illustrated in recent large-scale pretrained models such as SDXL~\cite{podell2023sdxl} for images and AudioLDM2~\cite{liu2023audioldm2} for audio, our empirical analysis shows that a fixed guidance scale is often suboptimal across diverse prompts and modalities: depending on the strength and complexity of the conditioning input, default values may lead to oversaturation, washed-out textures, or weak alignment in images, and to distorted timbre, artifacts, or weak semantic correspondence in audio.
For example, in SDXL~\cite{podell2023sdxl}, we observe that although the default parameter provides generally good global performance, it is not always optimal for specific prompts (see Fig.~\ref{fig:scale1} and Fig.~\ref{fig:scale2}). The best scale often depends on the complexity of the textual description and the detailed content information, while prompt-specific adjustments can significantly improve both fidelity and semantic alignment.

In this work, we take a first step toward revisiting guidance scale as an inference-time variable rather than a fixed hyperparameter. While one could naively perform iterative inference across multiple scales for each sample and retain the best output according to objective metrics, such an approach is prohibitively time-consuming and computationally inefficient, rendering it impractical in real-world scenarios. Instead, we demonstrate that a lightweight network (up to 4M) can reliably approximate the behavior of a pretrained diffusion model, thereby providing an effective surrogate for identifying the optimal guidance scale. We propose a prompt-aware framework that learns to predict scale-dependent generation quality from semantic and complexity representations of the input prompt. By integrating a Lightweight agent score estimator with pretrained diffusion models, our method enables per-prompt selection of the guidance scale without retraining or costly grid search. We validate the approach on both image (MSCOCO~2014 with SDXL)~\cite{lin2014microsoft} and audio (AudioCaps with AudioLDM2)~\cite{kim2019audiocaps} generation, demonstrating consistent improvements in fidelity, semantic alignment, and perceptual quality over vanilla CFG. These results highlight that guidance scale, though simple, can be re-examined as a powerful lever for improving pretrained diffusion backbones across modalities.

\begin{algorithm}[t]
\caption{Prompt-Aware Guidance}
\label{alg:promptaware}
\begin{algorithmic}[1]
\State \textbf{Input:} prompt set $\mathcal{P}$, scale set $\mathcal{S}$, pretrained parameters $\theta$
\State \textbf{Output:} generated image or audio sample $I$

\State \textbf{Training:}
\For{each $(p,\omega)$}
    \State Generate $I^{(j)}_{p,\omega} \sim p_\theta(\,\cdot \mid p,\omega\,)$ for $j=1,\ldots,N_g$
    \State Compute $q^{(j)}_m(p,\omega)$ for all $m \in \mathcal{M}$
    \State Aggregate $\mathbf{q}(p,\omega)=\tfrac{1}{N_g}\sum_{j=1}^{N_g} \big[q^{(j)}_m(p,\omega)\big]_{m\in\mathcal{M}}$
    \State Predict $\widehat{\mathbf{q}}(p,\omega)=g_\phi\!\big(\mathbf{h}(p,\omega)\big)$
    \State Update $\phi$ by minimizing $\big\|\widehat{\mathbf{q}}(p,\omega)-\mathbf{q}(p,\omega)\big\|_2^2$
\EndFor

\State \textbf{Inference:}
\For{each $p$}
    \State Evaluate $\widehat{\mathbf{q}}(p,\omega)$ for all $\omega\in\mathcal{S}$
    \State $\omega^\star(p)=\arg\max_{\omega\in\mathcal{S}}\!\left[\,\mathbf{w}^\top \widehat{\mathbf{q}}(p,\omega) - \alpha\,(\omega-\mu_\omega)^2\,\right]$
    \State Generate $I$ under $\omega^\star(p)$ by reversing diffusion:
    
           $I \sim \int \prod_{t=1}^{T} p_\theta\!\big(x_{t-1}\mid x_t,p,\omega^\star(p)\big)\,dx_{1:T}$
\EndFor
\end{algorithmic}
\end{algorithm}

\begin{figure*}[t]
  \centering
  \includegraphics[width=\textwidth,trim={20pt 10pt 20pt 10pt},clip]{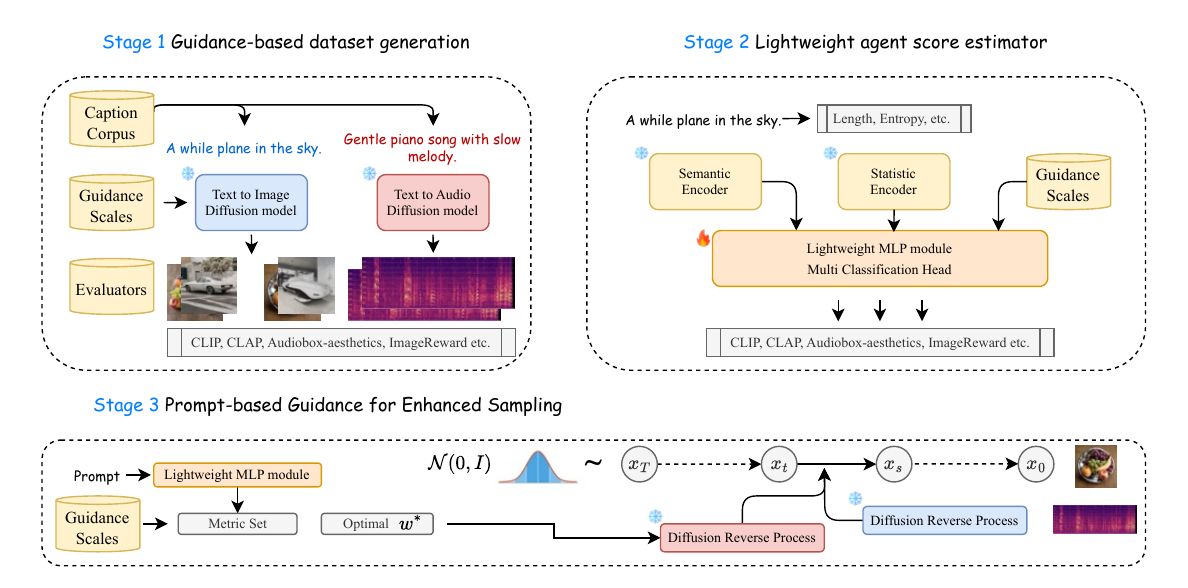}
  \captionsetup{font=small} 
  \caption{Overview of the proposed framework. Stage~1 constructs a guidance-based dataset by generating multi-scale samples from both text-to-image (SDXL) and text-to-audio (AudioLDM2) diffusion models. Each sample is evaluated with modality-specific metrics such as CLIP, ImageReward, or AudioBox-Aesthetics. Stage~2 trains a lightweight predictor that integrates semantic embeddings (CLIP/CLAP) with statistical complexity features (e.g., length, entropy). Stage~3 leverages the predictor at inference to select prompt-dependent guidance scales, which are then used in the diffusion reverse process for enhanced sampling.}
  \label{fig:model}
\end{figure*}

\section{Methodology}
\label{method}

We first revisit classifier-free guidance (CFG) and analyze its limitations under a fixed global scale, and then present our prompt-aware framework, which leverages scale-dependent supervision to train a lightweight predictor and adaptively select guidance strength at inference.

\subsubsection*{A. Baseline: Classifier-Free Guidance}

\textbf{Preliminary.}
Diffusion models~\cite{songscore,ho2020denoising} can be trained by denoising score matching, which is equivalent to maximizing the evidence lower bound (ELBO). Formally, the optimal denoising network $\epsilon_\theta$ converges to the score function of the perturbed data distribution:
\begin{equation}
\epsilon_\theta^\ast(x_t \mid p) = -\sigma_t \nabla_{x_t} \log p_t(x_t \mid p),
\tag{1}
\end{equation}
where $\sigma_t$ is the variance of Gaussian noise added at step $t$, and $p_t(x_t \mid p)$ denotes the marginal of the noised sample conditioned on prompt $p$.

At inference, generation proceeds by reversing the forward diffusion. Let $x_t$ denote the latent at step $t$. The one-step reverse transition is
\begin{equation}
p_\theta(x_{t-1} \mid x_t, p, \omega) 
= \mathcal{N}\!\big(x_{t-1}; \mu_\theta(x_t, p, \omega, t), \sigma_t^2 I \big),
\tag{2}
\end{equation}
where the mean $\mu_\theta$ is parameterized via the predicted noise. Unrolling all steps yields the distribution of the generated sample $I$:
\begin{equation}
p_\theta(I \mid p, \omega) 
= \int \prod_{t=1}^{T} p_\theta(x_{t-1} \mid x_t, p, \omega)\, dx_{1:T}.
\tag{3}
\end{equation}

Classifier-Free Guidance (CFG) modifies the noise estimate at each step by interpolating unconditional and conditional predictions:
\begin{equation}
\hat{\epsilon}_\theta(x_t, p, t; \omega) 
= (1-\omega)\,\epsilon_\theta(x_t, \emptyset, t) 
+ \omega\,\epsilon_\theta(x_t, p, t),
\tag{4}
\end{equation}
where $\epsilon_\theta(x_t, p, t)$ and $\epsilon_\theta(x_t, \emptyset, t)$ denote conditional and unconditional estimates, respectively, and $\omega$ is the guidance scale. From a distributional perspective, CFG shifts the sampling distribution to
\begin{equation}
p^s(x \mid p) \propto p(x \mid p) \left[ \frac{p(x \mid p)}{p(x)} \right]^s,
\tag{5}
\end{equation}
where $s$ corresponds to the guidance scale: larger $s$ enhances alignment but reduces diversity.

\subsubsection*{B. Overall Framework}

Our framework integrates scale-dependent supervision with prompt-based predictors to enable per-prompt guidance selection. As shown in Fig.~\ref{fig:model}, it consists of three stages:

\textbf{1) Guidance-based dataset generation.}  
For each prompt $p$, multiple generations are produced across candidate scales $\omega \in \mathcal{S}$. Each output is evaluated with modality-specific metrics (KID, CLIP, ImageReward~\cite{xu2023imagereward}, Precision/Recall for images; AudioBox-Aesthetics for audio). Averaged scores form the oracle label $\mathbf{q}(p,\omega)$.

\textbf{2) Lightweight agent score estimator.}  
Given a prompt $p$, we extract semantic and complexity embeddings:
\begin{equation}
\mathbf{e}(p) = E_{\text{text}}(p)\in\mathbb{R}^{d_e}, \qquad
\mathbf{c}(p) = W_c\,\mathbf{r}(p)+\mathbf{b}_c,
\tag{6}
\end{equation}
where $E_{\text{text}}$ is instantiated as CLIP~\cite{radford2021learning} for images or CLAP~\cite{elizalde2023clap} for audio, and $\mathbf{r}(p)$ encodes lexical/syntactic statistics (length, entropy, perplexity, modifier diversity, etc.).  
The joint representation is
\begin{equation}
\mathbf{h}(p,\omega) = \big[\,\mathbf{e}(p);\ \mathbf{c}(p);\ \omega\,\big].
\tag{7}
\end{equation}
A lightweight predictor then outputs estimated quality:
\begin{equation}
\widehat{\mathbf{q}}(p,\omega)=g_\phi(\mathbf{h}(p,\omega))\in\mathbb{R}^{d_q}.
\tag{8}
\end{equation}

\textbf{3) Prompt-based guidance for enhanced sampling.}  
At inference, the predictor evaluates candidate scales and selects the optimal $\omega^\star(p)$:
\begin{equation}
\omega^\star(p) = \arg\max_{\omega\in\mathcal{S}}
\Big[\,\mathbf{w}^\top \widehat{\mathbf{q}}(p,\omega) - \alpha(\omega-\mu_\omega)^2\,\Big],
\tag{9}
\end{equation}
where $\mathbf{w}\in\mathbb{R}^{d_q}_{\ge0}$ are task-specific weights, and the penalty term regularizes abnormal scales around an anchor $\mu_\omega$. The final sample is then generated under $\omega^\star(p)$ using the CFG-modulated reverse process:
\begin{equation}
p_\theta(I \mid p) = \int \prod_{t=1}^{T} 
p_\theta(x_{t-1}\mid x_t,p,\omega^\star(p))\,dx_{1:T}.
\tag{10}
\end{equation}
This ensures prompt-dependent guidance that balances fidelity, alignment, and perceptual quality.

The overall pipeline of our proposed framework is summarized in Algorithm~\ref{alg:promptaware}.

\section{Experiment}
\label{sec:pagestyle}

\subsubsection*{A. Model, Dataset, and Evaluation}
\label{sssec:dataset}

We adopt SDXL~1.0 for image generation and AudioLDM2~\cite{liu2023audioldm2} for audio generation as the backbone diffusion models.  
Both are large-scale pretrained models with strong zero-shot generalization, making them suitable testbeds for evaluating inference-time guidance strategies.  
Training data are constructed from MSCOCO~2014 (images) and AudioCaps (audio) by pairing sampled prompts with candidate guidance scales and generating multiple outputs for each pair.  
This procedure explores scale-dependent behaviors and yields a synthetic dataset where per-scale averages of evaluation scores serve as oracle supervision for training the lightweight predictor.

For images, training supervision is derived from four complementary metrics: KID, CLIP score, ImageReward, and Precision/Recall.  
KID is chosen as the primary signal due to its stability with limited samples, while CLIP score measures semantic alignment, ImageReward reflects human preference, and Precision/Recall capture the fidelity–diversity trade-off.  
For audio, training relies on AudioBox-Aesthetics~\cite{tjandra2025aes}, a perceptual proxy closely aligned with human judgments.

Validation uses larger, disjoint prompt sets: 3k from MSCOCO~2014 for images and 1k from AudioCaps for audio.  
For images, evaluation employs FID as the primary indicator of fidelity and CLIP score for semantic alignment, while for audio, AudioBox-Aesthetics serves as the sole perceptual metric.  
This setup ensures both efficient supervision and comprehensive cross-modal evaluation.

\begin{table}[t]
\centering
\caption{Comparison of Non-adaptive CFG and our prompt-aware method on images. }
\label{tab:results}
\begin{tabular}{|l|c|c|}
\hline
Method & FID$\downarrow$ & CLIP$\uparrow$ \\
\hline
No Guidance ($\omega=1.0$) & 62.44 & 0.27 \\
Non-adaptive ($\omega=w_\text{defalut}^{\text{image}}$) & 31.04 & 0.31 \\
Prompt-aware ($\omega^\star(p)$) & 30.74 & 0.33\\
\hline
\end{tabular}
\end{table}

\begin{table}[t]
\centering
\caption{Comparison of Non-adaptive CFG and our prompt-aware method on audios. }
\label{tab:aes_results}
\begin{tabular}{|l|c|c|c|c|}
\hline
Method & CE$\uparrow$ & CU$\uparrow$ & PC$\uparrow$ & PQ$\uparrow$ \\
\hline
No Guidance ($\omega=1.0$) & 3.62& 5.18& 3.13& 5.76\\
Non-adaptive ($\omega=w_\text{default}^{\text{audio}}$) & 3.66& 5.25& 3.04& 5.79\\
Prompt-aware ($\omega^\star(p)$) & 3.68& 5.22& 3.16& 5.81\\
\hline
\end{tabular}
\end{table}

\subsubsection*{B. Results and Analysis}
\label{sssec:results}

We compare the proposed prompt-aware guidance with two baselines: 
(i) \emph{No Guidance}, where classifier-free guidance is disabled, and 
(ii) \emph{Vanilla CFG}, where a single fixed scale is applied to all prompts. 
Table~\ref{tab:results} reports the quantitative results for the image setting on the validation set described in Sec.~\ref{sssec:dataset}, using FID as the primary measure of distributional fidelity and CLIP score for semantic alignment. 
For audio, Table~\ref{tab:aes_results} presents evaluation with AudioBox-Aesthetics, which directly reflects perceptual quality.
In addition, we conduct ablation studies on the training-time image generation design in Table~\ref{tab:ablation_image}.

Compared to the baselines, Vanilla CFG exhibits a trade-off: smaller scales tend to maintain diversity but reduce alignment, while larger scales improve alignment but can degrade fidelity.
Our method addresses this issue by selecting $\omega^\star(p)$ for each prompt, leading to more stable performance across prompts and mitigating the drawbacks of a fixed scale.

Furthermore, Table~\ref{tab:ablation_image} shows that relying solely on KID and CLIP cannot fully capture overall generation quality and may even yield controversial effects, as improvements in alignment do not necessarily translate to perceptual preference or distributional fidelity. By contrast, combining all four metrics: KID, FID, CLIP, and ImageReward, together with the quadratic regularization term provides a more comprehensive and consistent supervision signal, highlighting the importance of multi-metric integration for stable and reliable scale selection.

Beyond quantitative improvements, we also observe consistent trends across both visual and auditory modalities: 
the proposed framework adapts guidance strength in a prompt-specific manner, yielding sharper visual details and more coherent audio textures. 
Moreover, the predictor exhibits robustness under diverse prompt distributions, including long, compositional, or ambiguous queries, 
suggesting that the approach generalizes well beyond the training pool. 
Together, these findings highlight the practicality of prompt-aware guidance as a lightweight yet effective mechanism for enhancing pretrained diffusion models.


\begin{table}[t]
\centering
\caption{Ablation studies on image generation used for training. 
$\uparrow$ indicates higher is better}
\label{tab:ablation_image}
\begin{tabular}{|l|c|c|}
\hline
\textbf{Method} & \textbf{FID} $\downarrow$ & \textbf{CLIP} $\uparrow$ \\
\hline
Non-adaptive& 31.04& 0.31\\
\hline
KID and CLIP&  31.81&  0.31\\
\hline
All metrics   & 30.74 & 0.33 \\
\hline
\end{tabular}
\end{table}

\section{Conclusion}
\label{sec:conclusion}

We introduced a prompt-aware classifier-free guidance framework for diffusion models that selects guidance strength conditioned on the input prompt. By leveraging supervision across multiple scales, the proposed method learns to predict quality-aware utilities and thereby improves generation quality without costly search. Experimental results on MSCOCO~2014 (image) and AudioCaps (audio) show consistent gains over vanilla CFG in terms of FID, KID, CLIP score, ImageReward, and AudioBox-Aesthetics. Future work will explore extending this approach to other conditional generation tasks, and further analyzing its efficiency and robustness under noisy or ambiguous prompts. We also plan to investigate more principled criteria for balancing fidelity, diversity, and alignment, which may provide deeper insights into the role of prompt-aware guidance in diffusion models.

\bibliographystyle{IEEEbib}
\bibliography{strings,refs}

\end{document}